\documentclass[aps,prb,twocolumn,showpacs,superscriptaddress,10pt]{revtex4-1}
\usepackage{graphicx,amsmath}

\begin{document}

\title{High density limit of the two-dimensional electron liquid with Rashba spin-orbit coupling}

\author{Stefano Chesi}
\affiliation{Department of Physics, Purdue University, West Lafayette, IN 47907, USA}
\affiliation{Department of Physics, University of Basel, 4056 Basel, Switzerland}
\affiliation{Department of Physics, McGill University, Montreal, Quebec, Canada H3A 2T8}

\author{Gabriele F. Giuliani}
\affiliation{Department of Physics, Purdue University, West Lafayette, IN 47907, USA}

\date{\today}

\begin{abstract}
We discuss by analytic means the theory of the high-density limit of the unpolarized two-dimensional electron liquid in the presence of Rashba or Dresselhaus spin-orbit coupling. A generalization of the ring diagram expansion is performed. We find that in this regime the spin-orbit coupling leads to small changes of the exchange and correlation energy contributions, while modifying also, via repopulation of the momentum states, the non-interacting energy. As a result, the leading corrections to the chirality and total energy of the system stem from the Hartree-Fock contributions. The final results are found to be vanishing to lowest order in the spin-orbit coupling, in agreement with a general property valid to every order in the electron-electron interaction. We also show that recent quantum Monte Carlo data in the presence of Rashba spin-orbit coupling are well understood by neglecting corrections to the exchange-correlation energy, even at low density values.
\end{abstract}

\pacs{71.10.Ca, 71.10.-w, 75.70.Tj, 73.61.Ey}

\maketitle

\setlength\arraycolsep{0pt}

\section{INTRODUCTION}

The effect of the spin-orbit coupling in semiconductor heterostructures, and more specifically 
the role of the Rashba spin-orbit interaction induced by the asymmetry of the transverse 
confining potential in electronic two-dimensional systems,\cite{rashbaSO} has attracted in 
recent years great interest. While a main motivation lies in the potential of new 
applications,\cite{datta89,wolf01,zutic04} based on the control of the single-particle 
spin-dependent dynamics through electrical gating,\cite{rokhinson04,khodas04,chen05,meier07,chesi07c,Chesi2011qpc} 
the corresponding many-body problem is of fundamental relevance and not yet fully investigated.

The two-dimensional electron liquid, in the presence of Coulomb interaction and a rigid 
neutralizing background, is a classic problem in solid-state physics, if the simplest 
effective mass approximation is assumed for the kinetic term.\cite{TheBook} On the other 
side, concomitant band-structure effects have often large observable consequences.\cite{WinklerSpringer03,winkler05,Chesi2011qpc} 
The special form of a generalized spin-orbit coupling applicable in a number of cases is
described in Ref.~\onlinecite{Chesi2007exchange}, together with a detailed analysis of the exchange 
energy. Several other aspects of the electron-electron interaction in the presence of spin-orbit couplings were addressed in Refs.~\onlinecite{Chen1999,Saraga2005,Giuliani2005,Wang2005,Pletyukhov2006,Badalyan2009,Dimitrova2005,Schliemann2006,Schliemann2010,Chesi2007phasediagr,ChesiPhD,Ambrosetti2009,Nechaev2009,Nechaev2010,Chesi2010exact,Zak2010,Abedinpour2010,Zak2010,Agarwal2011}. For example the quasiparticle properties,\cite{Chen1999,Saraga2005,ChesiPhD,Nechaev2009,Nechaev2010,Chesi2010exact,Agarwal2011} the Hartree-Fock phase-diagram,\cite{Giuliani2005,Chesi2007phasediagr,ChesiPhD} the spin susceptibility,\cite{Dimitrova2005,ChesiPhD,Zak2010,Agarwal2011} and the plasmon modes.\cite{Wang2005,Pletyukhov2006,Badalyan2009,Schliemann2010,Agarwal2011}
We restrict ourselves here to pure Rashba,\cite{rashbaSO} or equivalently Dresselhaus,\cite{dresselhaus55,comment_dresselhaus} spin-orbit coupling. In this case, quantum Monte Carlo data for the total energy were recently obtained in Ref.~\onlinecite{Ambrosetti2009}. Numerical results for the total energy also appear in Ref.~\onlinecite{Abedinpour2010}, within the random phase approximation (RPA). We focus in the following on the high-density limit, when the effect of the Coulomb interaction can in general be studied perturbatively. 

The leading correction 
to the non-interacting energy is the exchange contribution, while higher order 
terms correspond to higher powers of the standard density parameter $r_s$ 
[see Ref.~\onlinecite{TheBook} and Eq.~(\ref{alpha_rs_def}) below]. The first two 
terms of the small $r_s$ expansion of the correlation energy in two-dimensions 
are well known in the absence of spin-orbit coupling.\cite{rajagopal77} 
They consist of the second-order correlation energy and a $r_s \ln r_s$ 
contribution which is obtained as an infinite sum of diverging ring diagrams. 
The elegant resummation procedure patterns the treatment of the leading $\ln r_s$ 
correlation energy in the three-dimensional case.\cite{gellmann57} An exact formula 
for the polarization dependence of the $r_s \ln r_s$ contribution was recently 
derived in Ref.~\onlinecite{chesi07a}.

Here, an additional dependence on the (dimensionless) Rashba coupling $\bar \alpha$ is present. 
However, the strength of the spin-orbit interaction is more appropriately expressed 
in terms of a parameter $g$ which is proportional to $r_s$ [see Ref.~\onlinecite{Chesi2007exchange} and Eq.~(\ref{gdef}) below]. Hence, an 
additional density dependence is introduced by $g$. This makes the effect of the 
spin-orbit interaction small, since the correction to the total exchange-correlation 
energy is multiplied by a factor at most of order $g^2$ (i.e. an even power in the 
spin-orbit coupling), which is vanishingly small at high density. Therefore, an accurate 
result is obtained perturbatively. In fact, a general argument for the energy expansion was derived in Refs.~\onlinecite{ChesiPhD,Chesi2010exact} and implies that the $g^2$ term is actually vanishing (see also Ref.~\onlinecite{Abedinpour2010}). 

In this paper, we analyze how the high-density expansion of the energy is modified in the presence of spin-orbit coupling.
The explicit analytic form of the leading exchange-correlation correction is obtained in the 
following and found indeed to be proportional to $g^4 \ln g$, from the exchange energy. The 
second-order correlation energy is studied numerically, as in the case without spin-orbit interaction, 
and the extension of the ring-diagram sum is also discussed, and shown to display a 
non-analytic behavior in the limit of small $r_s$ and $g$. Corrections to these higher 
order contributions are also found to be of higher order than $g^2$, as expected, and can therefore be usually neglected. Finally, while the main body of the paper is devoted to the asymptotic expansion at small $r_s$ (a regime often relevant for heterostructures with large spin-orbit coupling), we also analyze the quantum Monte Carlo results of Ref.~\onlinecite{Ambrosetti2009}, which are all at $r_s \geq 1$. We propose here a simple interpolation formula for the energy which is in remarkable agreement with the numerical data.

The detailed outline of the paper is as follows: in Sec. \ref{formulation_Section} we formulate 
the problem and establish our notation. We review the properties of the non-interacting system 
and the known results for the exchange-correlation energy in the absence of spin-orbit coupling. We 
also define here the corrections to the exchange-correlation energy of the electron liquid due 
to the spin-orbit interactions, which are the main focus of our work. We first show in Sec. \ref{diagrams_Section} 
that such corrections are generally small, by reminding the reader about some useful exact properties of the 
perturbative expansion,\cite{ChesiPhD,Chesi2010exact} and by an analysis of the quantum Monte Carlo results 
of Ref.~\onlinecite{Ambrosetti2009}. We then explicitly determine such corrections in the high-density limit of the electron gas. 
The exchange energy, the second-order correlation terms, and the classic ring expansion of Ref.~\onlinecite{rajagopal77} 
are revisited and extended in Sec.~\ref{Coulomb_Section} where we obtain the change of the exchange-correlation 
energy and of the momentum space occupation to leading order in the spin-orbit coupling. Both analytical and numerical results are provided, which 
are summarized in Sec. \ref{discussion}. In this last section, an alternate physical limit
is also discussed. Finally, the details of a number of calculations have been provided in 
Appendices \ref{deriv2direct}, \ref{deriv2ringdiagrams}, and \ref{different_limit}.

\section{\label{formulation_Section} FORMULATION OF THE PROBLEM}

The system is described by the hamiltonian
\begin{equation}
\label{Hint}
\hat {H} ~=~ \sum_i \hat H_{0}^{(i)} + \frac{1}{2}\sum_{i\neq j}
\frac{e^2}{| \hat{\bf r}_i- \hat{\bf r}_j|},
\end{equation}
where terms related to the presence of a uniform neutralizing background have been omitted
for simplicity. The single particle operator $\hat H_{0}$ is given by:
\begin{equation}
\label{H_0}
\hat{H}_{0} ~=~
\frac{ \hat{{\bf p}}^2}{2m}+\alpha \,(\hat{\sigma}_{x} \hat{p}_{y}- 
\hat{\sigma}_{y} \hat{p}_{x} ),
\end{equation}
where we consider electrons confined in the $(x,y)$ plane and $\hat{\sigma}_{x(y)}$ 
are Pauli matrices. The spin-orbit term is usually referred to as a linear Rashba 
spin-orbit coupling and is generally present when the confining potential in the $z$ direction of
a quantum well is asymmetric.\cite{rashbaSO} An equivalent term, the Dresselhaus spin-orbit 
coupling, arises instead for a lack of inversion symmetry in the crystal structure.\cite{dresselhaus55}
We consider Eq.~(\ref{H_0}) as a model case, although a similar analysis can be carried out for other types of spin-orbit interaction, relevant in other experimental cases.\cite{WinklerSpringer03,Chesi2007exchange,Chesi2011qpc}

We use in the following dimensionless units. The properties of the electron 
liquid are completely determined by
\begin{equation}\label{alpha_rs_def}
\bar\alpha=\frac{\hbar\alpha}{e^2}   \qquad {\rm and} \qquad r_s=\frac{1}{\sqrt{\pi n a_B^2}}, 
\end{equation}
where $\bar\alpha$ is a dimensionless spin orbit coupling and $r_s$ the usual density parameter, with
$a_B$ the effective Bohr radius and $n$ the number density. 
It is also useful to introduce the following dimensionless coefficient $g$
\begin{equation}\label{gdef}
g=\sqrt{2}\bar\alpha r_s .
\end{equation}
which better than $\bar\alpha$ represents the strength of the spin-orbit term.
In fact, $g$ is approximately equal to the ratio of the spin-orbit energy to the 
kinetic energy, which are respectively proportional to $g/r_s^2$ and $1/r_s^2$ [see also Eq. (\ref{nonint_energy})]. The wave vectors are expressed in terms of the Fermi wave vector $k_F=\sqrt{2 \pi n}$ 
and the energies are in $Ry$ units. Finally, throughout the
paper we often use the notation ${\bf p} = {\bf k} + {\bf q}$ and ${\bf p}^\prime = {\bf k}^\prime - {\bf q}$ (so that ${\bf q}$ will not 
explicitly appear in many expressions).

\subsection{Non-interacting electrons}

The non-interacting problem is completely determined by $g$. The eigenstates 
of $H_0$ can be written as
\begin{equation}
\label{phi0kpm}
\varphi_{ {\bf k} , \pm} ( {\bf r} )  =
\frac{e^{i {\bf k} \cdot {\bf r} }}{\sqrt{2L^2}}
\left(
\begin{array}{c}
\pm 1\\
i e^{i\phi_{\bf k}} 
\end{array}
\right)
\equiv
\frac{e^{i {\bf k} \cdot {\bf r} }}{\sqrt{L^2}}
| { \bf k } \pm \rangle ,
\end{equation}
where $L$ is the linear size of the system and $\phi_{\bf k}$ is the angle 
formed by ${\bf k}$ with the $x$ axis. The eigenstates have spin quantized 
perpendicular to the wave vector $\mathbf{k}$, as described by the spinor functions 
$| {\bf k} \pm \rangle$ [which are defined by Eq.~(\ref{phi0kpm})].  
The corresponding eigenenergies (in $Ry$ units) are equal to $\frac{2}{r_s^2} \, \epsilon_\pm(k)$ 
where
\begin{equation}
\label{epsilon0k2}
\epsilon_\pm(k) =  k^2  \-\mp g \, k   .
\end{equation} 

\begin{figure}
\includegraphics[width=0.4\textwidth]{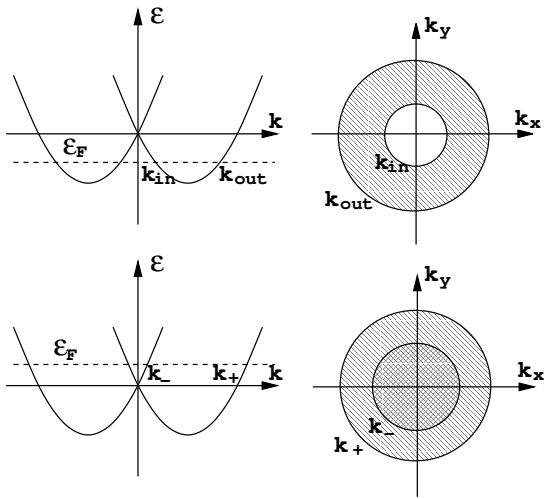}
\caption{\label{occupation} Two different ways of occupying non-interacting chiral
states
in $\mathbf{k}$ space.}
\end{figure}

We also define the generalized chirality $\chi$, which determines the occupation functions
$n_\pm(k)$ in momentum space.\cite{Giuliani2005,Chesi2007phasediagr} The two relevant regimes are depicted 
in Fig. \ref{occupation}. At high density (second panel of Fig. \ref{occupation}), two bands 
are occupied and $\chi$ coincides with the regular chirality: $\chi=\frac{N_+-N_-}{N_++N_-}$, where $N_\pm$ is the total number of electrons in each band.
The occupation takes the form
\begin{eqnarray}\label{chi}
n_\pm(k)=\theta(\sqrt{1\pm \chi}- k)     &   \quad {\rm for }~0 \leq\chi < 1   ,
\end{eqnarray}
where $\theta(x)$ is the usual step function.  
At low density (first panel of Fig. \ref{occupation}), 
the higher band is empty and the occupation function reads
\begin{eqnarray}\label{genchi}
n_+(k)=\theta(\sqrt{\chi+1}- k)-\theta(\sqrt{\chi-1}- k),
\end{eqnarray}
where $\chi \ge 1$. We note that Eq.~(\ref{genchi}) corresponds to a 
ring in momentum space and the regular chirality is 1 in this case, 
irrespectively of the precise form of the occupation. The Fermi surfaces are determined 
in all cases by the radii
\begin{equation}\label{kpm}
k_\pm=\sqrt{|1 \pm \chi|}.
\end{equation} 

The non-interacting energy at generic $\chi$ is expressed as follows 
(when $\chi<1$)
\begin{equation}\label{nonint_energy}
\mathcal{E}_0(g,r_s,\chi)=\frac{1+\chi^2}{r_s^2}-\frac{2g}{r_s^2}
\frac{\sqrt{|1+\chi|^3}-\sqrt{|1-\chi|^3}}{3}. 
\end{equation}
The first term is the kinetic energy, and one has to replace  $1+\chi^2$ with $2\chi$ if $\chi>1$. The second term is the spin-orbit
energy. The non-interacting ground state is specified by the value of $\chi$ which minimizes Eq.~(\ref{nonint_energy}) 
for given values of $\bar\alpha$ and $r_s$ and is therefore uniquely determined by the parameter $g$ of Eq.~(\ref{gdef})
\begin{eqnarray}\label{chi_nonint}
\chi_{0}(g)=
\left\{
\begin{array}{cl}
g \sqrt{1-\frac{g^2}{4}}    &\quad {\rm for}~0\leq g <\sqrt{2} ~,\\
\frac{g^2}{4}+\frac{1}{g^2} &\quad {\rm for}~g \geq \sqrt{2}.
\end{array}
\right.  
\end{eqnarray}
The corresponding ground state energy is obtained accordingly:
\begin{equation}\label{E_nonint}
\mathcal{E}_0(g,r_s)=\mathcal{E}_0(g,r_s,\chi_0(g)) ~.
\end{equation}

\subsection{Exchange-correlation energy}

The exchange-correlation energy of the electron liquid without spin-orbit coupling is a relatively 
well known quantity.\cite{TheBook} The perturbative expression
at high density reads\cite{rajagopal77} (for the unpolarized case)
\begin{equation}\label{energy_series}
\mathcal{E}_{xc}(r_s)=-\frac{8\sqrt{2}}{3\pi r_s}-0.385 
-\frac{2\sqrt{2}}{3\pi} (10-3\pi)\, r_s \ln r_s + \ldots  ~,
\end{equation}
where the first term is the exchange energy. The constant results from 
the numerical integration of the second order correlation energy,
and the last contribution is obtained from the infinite sum of diverging 
ring diagrams, similar to the original calculation for the three-dimensional case.\cite{gellmann57}
At generic values of the density, $\mathcal{E}_{xc}(r_s)$ is obtained
numerically with the Monte Carlo method\cite{attaccalite02,TheBook} 
(see however Ref.~\onlinecite{chesi07a} for the polarized case). 

On the other hand, the exchange-correlation correction due to the spin-orbit coupling 
is to date not accurately known. We introduce the following definition
\begin{equation}\label{deltaExc}
\mathcal{E}(g,r_s)=\mathcal{E}_0(g,r_s)+\mathcal{E}_{xc}(r_s)+\delta \mathcal{E}_{xc}(g,r_s) ~,
\end{equation}
where the non-interacting energy is given by Eq.~(\ref{E_nonint}) and 
$\delta \mathcal{E}_{xc}(g,r_s)$ represents the correction to the exchange-correlation energy 
associated with the spin-orbit coupling.
The latter is generally neglected, for example in density functional studies 
including spin-orbit interactions.\cite{valinrodriguez02a,valinrodriguez02b} 
A partial justification to this procedure is given in Ref.~\onlinecite{Chesi2010exact}, which shows that this correction is actually vanishing to quadratic order in $g$ for 
the particular case of the Rashba or Dresselhaus spin-orbit interaction. 
However, such a correction is not zero in general and can reasonably lead to 
important effects in the case of large $g$ (e.g., at low density) or for 
other types of spin-orbit interaction.\cite{Chesi2007exchange}

Formally, under the assumption that the system behaves as a Fermi liquid, the 
total energy of the interacting system can be obtained as a perturbative
expansion (see next section) constructed from a particular non-interacting state, 
as for example the one used to obtain Eq.~(\ref{nonint_energy}). Therefore, in the case with 
spin-orbit coupling, the total energy acquires an additional dependence from 
the chirality $\chi$ of the non-interacting state used in the perturbative expansion. 
This does not need to be the starting non-interacting ground state. 
As a consequence, we can quite generally write the total energy as
\begin{equation}\label{int_energy_chi}
\mathcal{E}(g,r_s,\chi)=\mathcal{E}_0(g,r_s,\chi)+
\mathcal{E}_{xc}(r_s)+\delta \mathcal{E}_{xc}(g,r_s,\chi) ~, 
\end{equation}
where $\mathcal{E}_0(g,r_s,\chi)$ is given by (\ref{nonint_energy}). The dependence of
the total energy on $\chi$, at given values of $g$ and $r_s$, is also obtained
in the Monte Carlo study of Ref.~\onlinecite{Ambrosetti2009}, where $\chi$ corresponds to the occupation of the
initial trial wave function. The data are reproduced in Fig.~\ref{MC_figure}.

The actual value of the interacting generalized chirality $\chi(g,r_s)$ is obtained by minimization of 
Eq.~(\ref{int_energy_chi}), which also yields the corresponding ground state energy
Eq.~(\ref{deltaExc}). It is important to realize that there are the two different contributions to 
$\delta \mathcal{E}_{xc}(g,r_s)$. The first one is given directly by 
$\delta \mathcal{E}_{xc}(g,r_s,\chi(g,r_s))$, while the second one arises from the 
renormalization of $\chi$ (i.e., the repopulation) in the non-interacting 
energy $\mathcal{E}_0(g,r_s,\chi)$.

\section{FORMAL PROPERTIES OF THE DIAGRAMMATIC EXPANSION}\label{diagrams_Section} 

We report here, in view of their usefulness, two exact results concerning the perturbative 
expansion of the energy and the quasiparticle self-energy. These results have been obtained
in Refs.~\onlinecite{ChesiPhD,Chesi2010exact} for a generic two-body potential $v(q)$. The first of the two 
results pertains to all diagrams $D$ contributing to the total energy and reads:
\begin{equation}\label{derivatives_rel_D}
\left.\frac{\partial^2 D}{\partial^2\chi}\right|_0 =
\left.\frac{\partial^2 D}{\partial^2 g}\right|_0 =
-\left.\frac{\partial^2 D}{\partial \chi \partial g}\right|_0  ~.
\end{equation}
This allows one to infer that, for small $g$ and $\chi$, the total correction to 
the exchange and correlation energy must behave like
\begin{equation}\label{diagr_generic_form}
\delta \mathcal{E}_{xc}(g,\chi) = C (g-\chi)^2 + \ldots  ~,
\end{equation}
where in general $C$ is an unknown constant.

The second result concerns the self-energy $\Sigma_\mu(k,\omega)$ which is seen to satisfy 
a similar exact relation to linear order in $g$ 
\begin{equation}\label{self-energy}
\left. \frac{\partial \Sigma_\mu(k,\omega)}{\partial g}\right|_0 =
-\frac{\mu}{2}  \frac{\partial \Sigma_0(k,\omega)}{\partial k}  ~,
\end{equation}
where $\Sigma_0(k,\omega)$ is the (interacting) self energy in the absence of
spin-orbit coupling (i.e. $g=\chi=0$). This relationship allows us to write:
\begin{equation}\label{self-energy2}
\Sigma_\mu(k,\omega) =
\Sigma_0(k-\frac{\mu g}{2},\omega)+ \ldots ~.
\end{equation}
From this formula we conclude that to linear order in $g$ all quasiparticles 
properties (e.g., the lifetime\cite{Saraga2005,Nechaev2009,Nechaev2010}) on the Fermi surfaces 
$k_\pm \simeq 1\pm \frac{g}{2}$ are the same as in the absence of spin-orbit coupling.

\begin{figure}
\begin{center}
\includegraphics[width=0.43\textwidth]{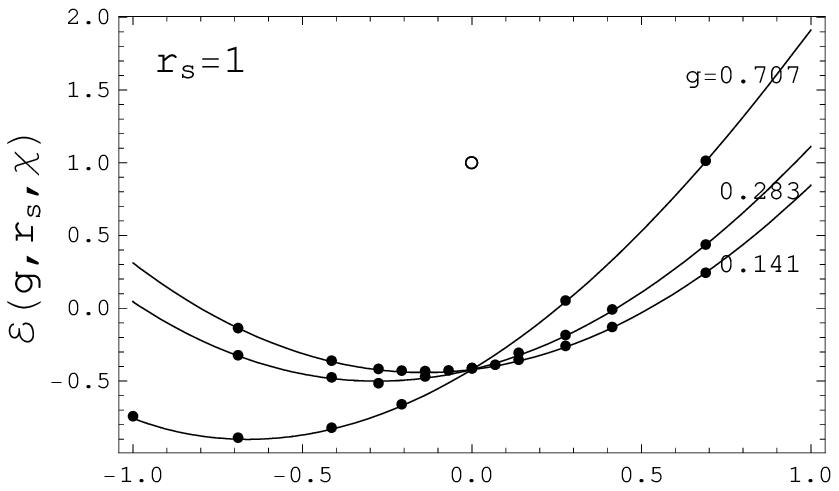}
\includegraphics[width=0.43\textwidth]{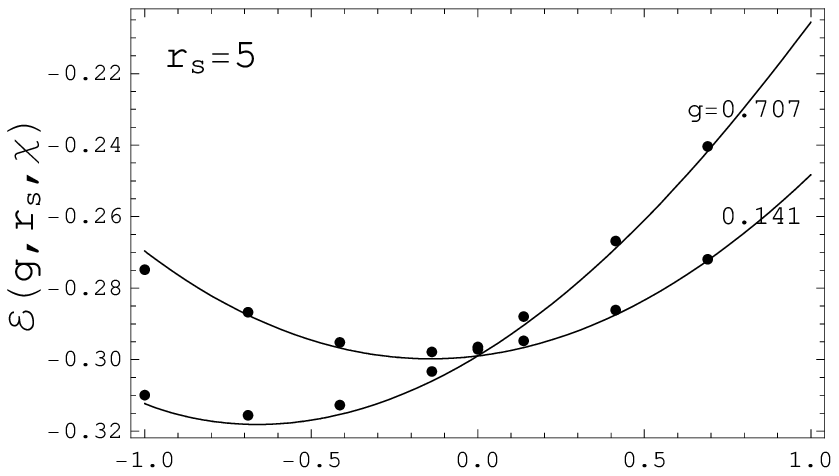}
\includegraphics[width=0.43\textwidth]{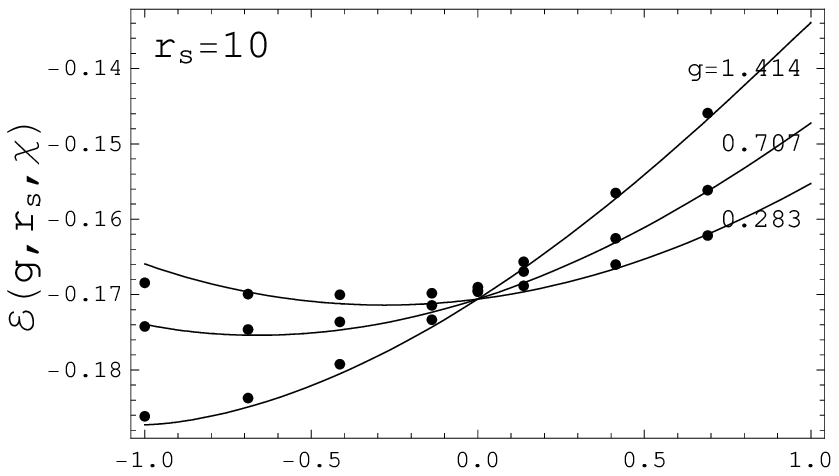}
\includegraphics[width=0.43\textwidth]{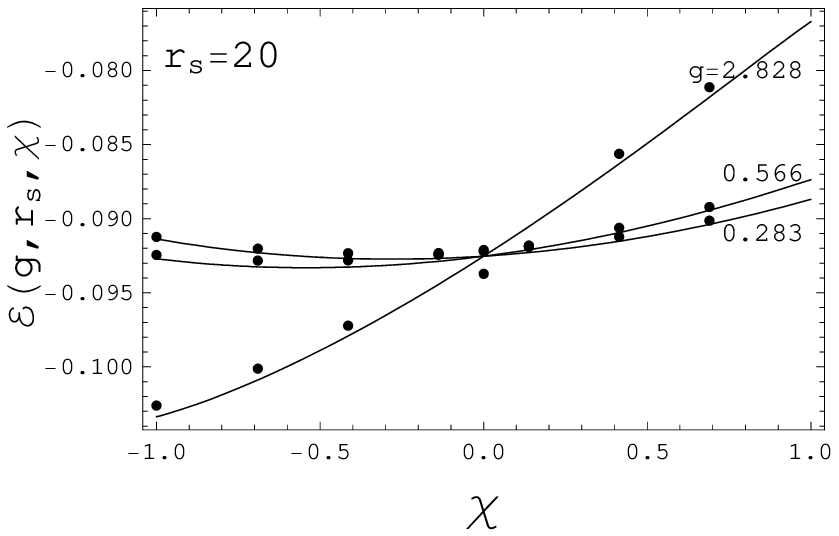}
\caption{\label{MC_figure} Numerical data (solid dots) from Ref.~\onlinecite{Ambrosetti2009} for the total energy $\mathcal{E}(g,r_s,\chi)$ of Eq.~(\ref{int_energy_chi}), as functions of $\chi$ and at different values of $g$ and $r_s$. We obtain the solid lines by setting $\delta\mathcal{E}_{xc}(g,r_s,\chi)=0$ in Eq.~(\ref{int_energy_chi}) and using the value of $\mathcal{E}_{xc}(r_s)$ of Ref.~\onlinecite{attaccalite02}. For reference, the empty dot in the top panel is the noninteracting energy without spin-orbit coupling (at $g=\chi=0$ and $r_s=1$). }
\end{center}
\end{figure}

For applications of these exact results we refer to Ref.~\onlinecite{Saraga2005}, which shows the validity
of Eq.~(\ref{self-energy2}) within the RPA, and to the numerical calculations of the
quasiparticle lifetime of Refs.~\onlinecite{Nechaev2009,Nechaev2010}. The total energy was obtained 
in the quantum Monte Carlo study of Ref.~\onlinecite{Ambrosetti2009} and we present in 
Fig.~\ref{MC_figure} their numerical data, together with the simple approximation of setting 
$\delta\mathcal{E}_{xc}(g,r_s,\chi)=0$ in Eq.~(\ref{int_energy_chi}). The value of the exchange-correlation 
energy $\mathcal{E}_{xc}(r_s)$ in the absence of spin-orbit coupling is taken from Ref.~\onlinecite{attaccalite02}. 
As seen, not only are the data at higher density ($r_s=1$) in very good agreement with
the curves obtained neglecting $\delta\mathcal{E}_{xc}(g,r_s,\chi)$, but also the low-density data are well described by this approximation.
This implies that the spin-orbit coupling shift of the exchange-correlation energy is generally small.
Notice that the accuracy of the numerical data does not allow us to extract the 
constant $C$ of Eq.~(\ref{diagr_generic_form}) as function of $r_s$. This we were able to obtain analytically in the limit 
of small $r_s$, as described in the following sections.

\section{PERTURBATIVE CONTRIBUTIONS OF THE COULOMB INTERACTION}\label{Coulomb_Section}

We examine in this section how the first terms of the high-density expansion are modified by 
the spin-orbit coupling. In particular, we discuss the exchange energy, the second-order 
correlation energy, and the sum of the ring diagrams. One has to notice that, 
if the bare value of the spin-orbit coupling $\bar\alpha$ 
is kept constant, the $r_s \to 0$ limit also corresponds to a vanishing strength 
of the spin orbit coupling $g$. 
This is clear from Eq.~(\ref{gdef}) and is simply understood as follows: at high density the spin-orbit energy grows like $\alpha \hbar k_F$, but becomes negligible with respect to the kinetic energy, which is proportional
to $k_F^2$. 

Furthermore, since $\chi(g,r_s)$ is given in first approximation by the non-interacting 
expression Eq.~(\ref{chi_nonint}), this limit corresponds also to a vanishing value 
of $\chi\simeq g$. Therefore, we will obtain an expansion of Eq.~(\ref{int_energy_chi}) 
in the small parameters $r_s$, $g$ and $\chi$, which are all of $O(r_s)$.

Using the result of the previous section at small $g$ and $\chi$ one can infer that 
to a generic contribution of order $\mathcal{O}(r_s^n)$ (in the absence of spin-orbit coupling)
corresponds a correction $\propto r_s^n (g-\chi)^2$ which is vanishing to lowest order 
for the ground-state energy. Therefore, the leading analytic contribution to $\delta\mathcal{E}_{xc}(g,r_s)$ in Eq.~(\ref{deltaExc}) is 
$\mathcal{O}(r_s^n g^4)= \mathcal{O}\left(r_s^{n+4}\bar\alpha^4 \right)$. This argument is 
valid for the exchange energy and the second-order correlation terms. On the other hand, 
due to their non-perturbative resummation of all orders, the series of diverging ring 
diagrams requires a more careful analysis.

\subsection{Exchange energy}

\begin{figure}
\begin{center}
\includegraphics[width=0.43\textwidth]{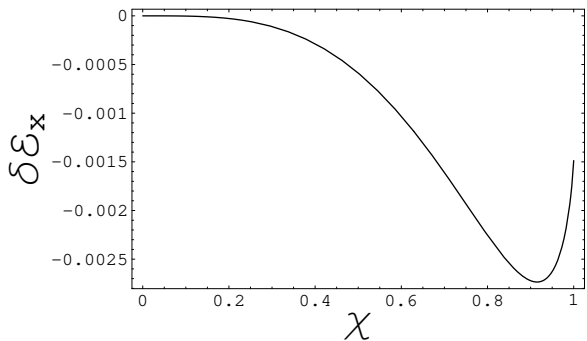}
\caption{\label{exchangeplot} Correction to the exchange energy (in $Ry$ units at $r_s=1$) 
for the unpolarized state with Rashba spin-orbit coupling, at finite values of $\chi$. To obtain
the value for generic $r_s$ one simply divides by $r_s$ [see Eq.~(\ref{exc_energy_chi})].  }
\end{center}
\end{figure}

The exchange energy is the main contribution of the electron interaction at high density.
The Hartree-Fock approximation of the two-dimensional electron liquid 
in the presence of spin-orbit coupling was already studied in our previous work.\cite{Chesi2007exchange,Chesi2007phasediagr,ChesiPhD} 
We derive here the explicit form of the exchange correction for the
specific case of Rashba or Dresselhaus spin-orbit coupling. 
This is expressed as follows\cite{Chesi2007exchange}
\begin{equation}\label{exc_energy_chi}
\delta \mathcal{E}^{(x)}_{xc}(g,r_s) ~=~
\frac{\delta \mathcal{E}_{x}(\chi)}{r_s} ~~, 
\end{equation}
where $\delta \mathcal{E}_{x}(\chi)$ is plotted in Fig. \ref{exchangeplot}. The expansion of $\delta \mathcal{E}_{x}(\chi)$ 
at small $\chi$ is given by
\begin{equation}\label{deltaEx}
\delta \mathcal{E}_{x}(\chi)=\frac{\sqrt{2}}{48\pi} \chi^4\left(\ln\frac{\chi}{8}
+\frac{23}{12}\right)+\ldots ~,
\end{equation}
where higher order terms are $\mathcal{O}(\chi^6)$. The term quadratic in $\chi$ 
is missing, in agreement with Eq.~(\ref{diagr_generic_form}) (with $C=0$).

By making use of Eqs.~(\ref{nonint_energy}), (\ref{deltaExc}), (\ref{exc_energy_chi}),
and (\ref{deltaEx}), one obtains that the energy is minimized when
\begin{equation}\label{deltachi}
\chi(g,r_s)=\chi_0(g)\left[
1-\frac{\sqrt{2}}{24 \pi}\, r_s g^2 \left(\ln \frac{g}{8}
+\frac{13}{6} \right)+ \ldots \right] ~,
\end{equation}
an expression that represents the analytic form of the small enhancement of the chirality 
numerically obtained in Ref.~\onlinecite{Chesi2007exchange}. The relative correction is of order 
$\mathcal{O}(r_s^3)$ and, as anticipated, it is quadratic in the spin-orbit coupling.

In calculating the correction to the total energy, one has to notice that $\chi_0(g)$
is a stationary value of the non-interacting energy, and therefore the corrections to
the non-interacting energy due to the renormalized value of $\chi$ are of order
$\mathcal{O}\left(\frac{( g^3 r_s )^2}{rs^2}\right)=\mathcal{O}(r_s^6)$ and can be neglected. 
The leading term is therefore given by
\begin{equation}\label{deltaExcResult}
\delta \mathcal{E}^{(x)}_{xc}(g,r_s) =
\frac{\sqrt{2} g^4}{48 \pi r_s} \left(\ln \frac{g}{8}+\frac{23}{12}\right)+ \ldots  ~,
\end{equation}
which represents the leading contribution to the exchange-correlation energy correction.
We discuss next the higher order correlation terms.

\subsection{Second-order correlation energy}

The second-order correlation energy $\mathcal{E}_2 (g,\chi)$ is obtained by standard perturbation theory.
In the intermediate state two electron-hole pairs are present, such that occupied 
states with wave vectors ${\bf k}$, ${\bf k}'$ and chiral indexes $\mu$, $\mu'$ have scattered to new
unoccupied states
\begin{eqnarray} \label{scattering}
( {\bf k} , \mu)\to ({ \bf p },  \nu)  
\quad {\rm and} \quad
( {\bf k}' , \mu') \to ({ \bf p }', \nu')  ,  \quad
\end{eqnarray} 
where ${\bf p} = {\bf k} + {\bf q}$ and ${\bf p}' = {\bf k}' - {\bf q}$. As it is well known, there are 
two different ways to scatter back to the original states. For direct processes
\begin{equation} 
( {\bf p} , \nu) \to ( {\bf k},  \mu)  
\quad {\rm and} \quad 
({ \bf p }', \nu')  \to  ( {\bf k}' , \mu') ,
\end{equation} 
which gives
\begin{eqnarray}\label{f2Ddef}
\mathcal{E}^D_2(g,\chi && )= -\frac{1}{4\pi^3}\sum_{\mu,\mu',\nu,\nu'}
\int \frac{d{\bf q}}{q^2}
\int d{\bf k}  
\int  d{\bf k}'  \nonumber \\
\times &&
\frac{
n_\mu (k) n_{\mu'}(k')
(1-n_{\nu} (p))
(1- n_{\nu '} (p'))
}{
\epsilon_{\nu}(p) - \epsilon_\mu (k) +
\epsilon_{\nu '}(p') - \epsilon_{ \mu '} (k')
} \nonumber \\
\times &&
|\langle {\bf p} \, \nu | {\bf k} \mu \rangle|^2 \,
|\langle {\bf p}'\, \nu' | {\bf k}' \mu' \rangle |^2 
.
\end{eqnarray}
For exchange processes 
\begin{equation} 
( {\bf p} , \nu)\to ({ \bf k }',  \mu')  
\quad {\rm and} \quad 
( { \bf p }', \nu')  \to  ( {\bf k} , \mu),
\end{equation} 
which corresponds to
\begin{eqnarray}
\label{f2Xdef}
\mathcal{E}^X_2(g, \chi &&  )= \frac{1}{4\pi^3}\sum_{\mu,\mu',\nu,\nu'}
\int \frac{d{\bf q}}{q  |{\bf k} - {\bf k}' + {\bf q}|}
\int d{\bf k}  
\int  d{\bf k}' \nonumber \\
\times && 
\frac{
n_\mu (k) n_{\mu'}(k')
(1-n_{\nu} (p))
(1- n_{\nu '} (p'))
}{
\epsilon_{\nu}(p) - \epsilon_\mu (k) +
\epsilon_{\nu '}(p') - \epsilon_{ \mu '} (k')
}  \nonumber \\
\times && 
\langle {\bf p} \, \nu | {\bf k} \mu \rangle  \, 
\langle {\bf k} \mu | {\bf p}' \nu ' \rangle \,
\langle {\bf p}'  \nu ' | {\bf k}' \mu ' \rangle \,
\langle {\bf k}' \mu' | {\bf p} \, \nu \rangle . ~~
\end{eqnarray}
Finally, the total second-order correlation energy is
\begin{equation}
\mathcal{E}_2(g,\chi)=\mathcal{E}^D_2(g,\chi)+\mathcal{E}^X_2(g,\chi)  ~.
\end{equation}

\begin{figure}
\begin{center}
\includegraphics[width=0.43\textwidth]{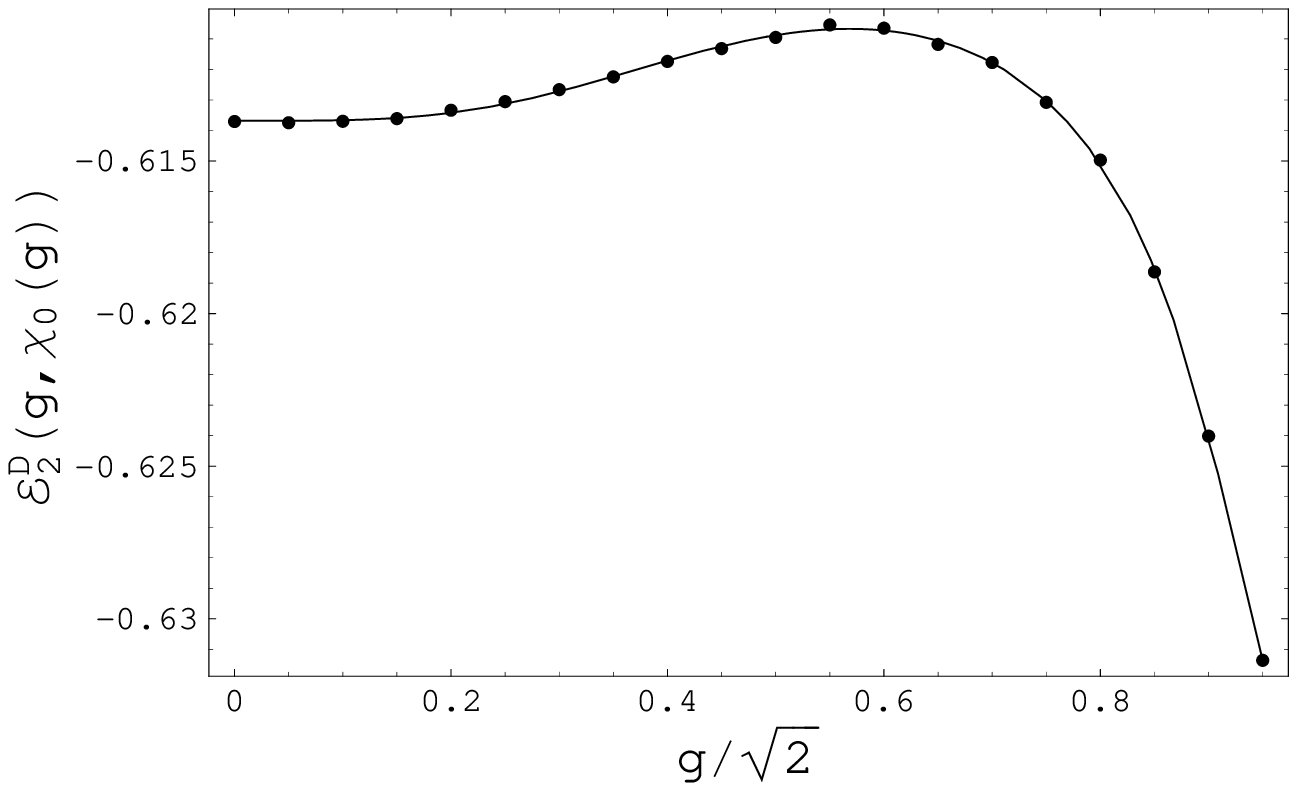}
\includegraphics[width=0.43\textwidth]{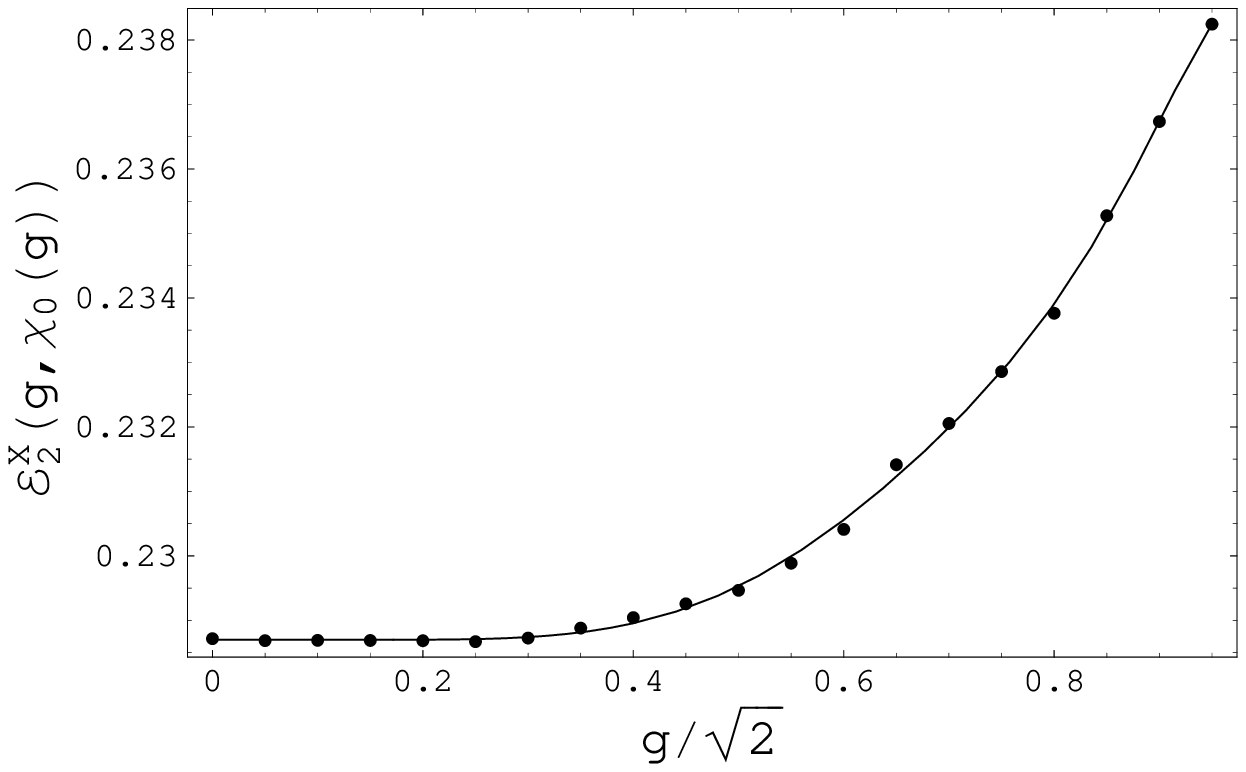}
\caption{\label{corr2} 
Plot of the second-order correlation energy for the non-interacting
ground state, as function of $g$. The generalized chirality is $\chi=\chi_0(g)$
and the range of both plots is such that $g<\sqrt{2}$, which gives $\chi<1$. 
The top panel shows the direct term $\mathcal{E}_2^D(g,\chi_0(g))$ and the lower panel the exchange term 
$\mathcal{E}_2^X(g,\chi_0(g))$. The points represent numerical results from Monte-Carlo
integrations of Eqs.~(\ref{f2Ddef}) and  (\ref{f2Xdef}), and the solid lines serve as a guide for the eye.
}
\end{center}
\end{figure}

\begin{figure}
\begin{center}
\includegraphics[width=0.43\textwidth]{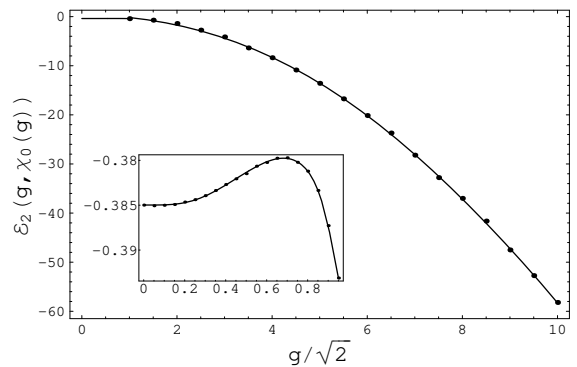}
\caption{\label{corr2sum} 
Plot of the total second-order correlation energy $\mathcal{E}_2(g,\chi_0(g))$,
with non-interacting occupation, as a function of $g$. The inset 
shows the region of small $g$, when $\chi<1$, and is obtained as the sum
of the curves displayed in the previous Fig. \ref{corr2}.
The points represent numerical results, 
and the solid lines serve as a guide for the eye.
}
\end{center}
\end{figure}

As in the case without spin-orbit coupling, Eqs.~(\ref{f2Ddef}) and (\ref{f2Xdef}) cannot be evaluated analytically in general. 
Furthermore, computing these multi-dimensional integrals with spin-orbit coupling is complicated  by 
the presence of singularities in the integration domain, from the energy denominators. In fact, 
the excitation energy is guaranteed to be positive when $\chi=\chi_{0}(g)$ but in the general case 
the energy denominator can be zero or negative. By restricting ourselves here to the simplest case $\chi=\chi_{0}(g)$,
which is correct to leading order at high density, we plot in Fig.~\ref{corr2} the direct and exchange 
second-order integrals as functions of $g$ for $g < \sqrt{2} $ (corresponding to $\chi <1$).
The sum of the two is plotted in a wider range of values in Fig.~\ref{corr2sum}. 
We notice in Fig.~\ref{corr2} that in the limit $g \to 0$ both functions $\mathcal{E}_2^D$
and $\mathcal{E}_2^X$ display a flat behavior in agreement with the vanishing of the
the $g^2$ contribution. The direct term is larger and dominates
the sum as displayed in the inset of Fig. \ref{corr2sum}. The characteristic behavior, 
similar to the case of the exchange energy (see Fig. \ref{exchangeplot}), suggests a $g^4 \ln g$
leading term. It is also remarkable that at large values of $g$ the correlation energy diverges. 
This limit of large spin-orbit coupling is highly non-perturbative, as
already revealed by the Hartree-Fock treatment. Within that approximation, the non-interacting 
states are strongly distorted by the Coulomb interaction and form special spin-textures 
in momentum space.\cite{Giuliani2005,Chesi2007phasediagr,ChesiPhD}

Analytic formulas for the second-order correlation energy can be obtained at small $g$ and $\chi$. Expanding to second order yields
\begin{eqnarray}
\label{E2D_final}
&\mathcal{E}^D_2(g,\chi)  \simeq - & 0.614 -\frac{(g-\chi)^2}{4} +\ldots ~, \\
\label{E2X_final}
&\mathcal{E}^X_2(g,\chi)  \simeq\quad   & 0.229+\ldots ~, 
\end{eqnarray} 
in agreement with Eq.~(\ref{diagr_generic_form}) (with $C=-\frac14$ and $C=0$ respectively).
The explicit calculation is detailed in Appendix~\ref{deriv2direct}.

\subsection{Ring diagrams}
The higher order terms ($n \geq 3$) in the perturbative treatment of the 
two-dimensional electron liquid are in general diverging for the
bare Coulomb interaction. However, a method to obtain the next leading correction 
to the correlation energy was devised for the three-dimensional 
case.\cite{gellmann57} It consists in summing to infinite order the (regularized) most diverging 
diagrams so that the final result is finite.\cite{TheBook} This method 
was applied in the two-dimensional case in Ref.~\onlinecite{rajagopal77} and is extended 
here with suitable modifications to include the Rashba or Dresselhaus spin-orbit coupling.

The expression of the ring diagrams reads ($n>1$)
\begin{equation}\label{rings_def}
\mathcal{E}^{(n)}_R(g,r_s,\chi)=
-\frac{(-1)^n}{\pi n r_s^2}
\int_{-\infty}^{+\infty} du
\int_0^\infty q^2 dq
\left(
\frac{Q_q(u)r_s}{2\sqrt{2}\pi q}
\right)^n ,
\end{equation}
where $Q_q(u)$ is given by
\begin{equation} \label{Qdef}
\sum_{\mu, \nu} 
\int d {\bf k} \, 
\, 
n_\mu(k)(1-n_\nu(p)) 
\frac{
(\epsilon_\nu(p)
-\epsilon_\mu(k))
|\langle {\bf p}\nu|
 {\bf k}\mu \rangle |^2}{
(\epsilon_\nu(p)
-\epsilon_\mu(k))^2/4+ u^2 q^2
} .
\end{equation}
$\mathcal{E}^{(n)}_R(g,r_s,\chi)$ has a (formal) dependence on $r_s^{n-2}$. Except for the
$n=2$ term, which is merely a compact formula for Eq.~(\ref{f2Ddef}), these expressions 
diverge. Summing them up to infinite order we arrive to
\begin{eqnarray}\label{rings_sum}
\mathcal{E}_R(g,r_s,\chi)=
\frac{1}{\pi r_s^2}
\int_{-\infty}^{+\infty} du
\int_0^\infty q^2 dq
\Bigg[
\ln \left(
1+\frac{Q_q(u)r_s}{2\sqrt{2}\pi q}
\right)
 \nonumber\\
-\frac{Q_q(u)r_s}{2\sqrt{2}\pi q} 
+\frac12 
\left( 
\frac{Q_q(u)r_s}{2\sqrt{2}\pi q}
\right)^2
\Bigg]. \qquad\qquad 
\end{eqnarray}

Rather than plunge into a numerical analysis, we endeavor next to extract the analytic 
behavior of this contribution on the variables $r_s$, $g$ and $\chi$. 
We begin by assessing the behavior of the function $Q_q(u)$ for $g, \chi\to 0$. 
We first notice that the $q \to 0$ limit of this function is not analytic. To see this
consider that, as shown in Appendix \ref{deriv2ringdiagrams}, for fixed $q$ 
and $g, \chi\to 0$, $Q_q(u)$ behaves accordingly to the general form given by 
Eq.~(\ref{diagr_generic_form}), i.e., like
\begin{equation} \label{Q_right_form}
Q_q(u) ~\simeq~ Q^{(0)}_q(u) ~+~ \frac{Q^{gg}_q(u)}{2} (g-\chi)^2 ~+~ \ldots 
\end{equation}
where $Q^{(0)}_q(u)$ is the value of $Q_q(u)$ for $g=\chi=0$ and 
$Q^{gg}_q(u) = \left. \frac{\partial^2 Q_q(u)}{\partial g^2}\right|_0$ is 
given in Eq.~(\ref{ddQ_gg_final}).
On the other hand this relationship does not hold as $q \to 0$ for fixed $g$ and $\chi$. In this
case one can neglect in Eq.~(\ref{Qdef}) terms involving scattering to the opposite branch. This
is justified since the factor $|\langle \nu {\bf p}|\mu \, {\bf k}\rangle |^2$ is $\sim 1$
for the intra-band and $\sim q^2$ for the inter-band contributions. Then
\begin{eqnarray} \label{Qresult}
Q_0(u)
\simeq 
\sum_{\mu}
\int \frac{d {\bf k}}{q} \, 
\, 
\frac{
2 \, n_\mu(k)(1-n_\mu(p)) \,
(k- \frac{\mu g}{2}) \cos\phi_{\bf k}
}{
(k-\frac{\mu g}{2})^2 \cos^2\phi_{\bf k} 
+ u^2
} \nonumber \\
=
\frac{2\pi k_+}{\tilde k_+ } R(\frac{u}{\tilde k_+}) +
\frac{2\pi k_-}{\tilde k_- } R( \frac{u}{\tilde k_- })
, \qquad
\end{eqnarray}
where $R(u)=1-1/\sqrt{1+1/u^2}$ and we have assumed $\chi<1$ 
so that $k_\pm=\sqrt{1 \pm \chi}$. We have also defined 
$\tilde k_\pm=k_\pm \mp g/2 $. 
In particular, by setting $\chi=g$, we obtain
\begin{align} \label{anomalous_Q}
Q_0(u)\simeq
4\pi R(u) - \frac{\pi u g^2}{2 (1+u^2)^{3/2}} ,
\end{align}
which of course is in violation of Eq.~(\ref{diagr_generic_form}).\cite{comment_no_bbroblem}

As a consequence one cannot immediately infer that Eq.~(\ref{rings_sum}) does in fact satisfy 
the general property Eq.~(\ref{diagr_generic_form}). This however proves not to be a problem since Eq.~(\ref{Qresult}) only
applies within a region of $q$ values of extension much smaller than $g$ and $\chi$. 
Accordingly, in order to obtain the leading contributions to $\mathcal{E}_R(g,r_s,\chi)$ 
when $\chi,g \to 0$ one can safely make use of Eq.~(\ref{Q_right_form}) 
for $Q_q(u)$. Therefore, the correction to the generic ring diagram can be formally written as
\begin{eqnarray}\label{d_rings_def}
\delta\mathcal{ E}^{(n)}_R=
-\frac{(-1)^n}{2\pi r_s^2}(g-\chi)^2 \int_{-\infty}^{+\infty} du 
\int_0^\infty q^2 dq\hspace{1cm} \nonumber \\
\times \left( \frac{Q^{(0)}_q(u) r_s}{2\sqrt{2}\pi q}
\right)^{n}\frac{Q^{gg}_q(u)}{Q^{(0)}_q(u)} , \qquad
\end{eqnarray}
to lowest order in  $g$ and $\chi$. 
Summing the geometric series (for $n>2$) we have
\begin{equation}\label{d_rings_sum}
\delta \mathcal{E}_R=
\frac{r_s}{16\pi^3}(g-\chi)^2 \int_{-\infty}^{+\infty} du 
\int_0^\infty dq \frac{\left[ Q^{(0)}_q(u) \right]^2 Q^{gg}_q(u)}
{2\sqrt{2}\pi q + Q^{(0)}_q(u)r_s} , \qquad
\end{equation}
where the integral is is logarithmically divergent at $r_s\to 0$, due to the small $q$ 
integration region. Therefore we can approximate 
\begin{equation}\label{d_rings_sum_approx}
\delta \mathcal{E}_R \simeq
-\frac{r_s\ln r_s }{2\sqrt{2}(2\pi)^4}(g-\chi)^2 
\int_{-\infty}^{+\infty} \left[ Q^{(0)}_0(u) \right]^2 Q^{gg}_0(u) du  , \qquad
\end{equation}
where $Q^{(0)}_0(u) \simeq 4\pi R(u)$ [with $R(u)$ defined below Eq.~(\ref{Qresult})]
and $Q^{gg}_0(u) \simeq -3\pi |u|/(1+u^2)^{5/2}$ which results from the small 
$q$ limit of (\ref{ddQ_gg_final}). 
We can then integrate Eq.~(\ref{d_rings_sum_approx}) to obtain
\begin{equation}\label{delta_ER_normal}
\delta \mathcal{E}_R(g,r_s,\chi) = \left(
\frac{56-15\pi}{40\sqrt{2}\pi}(g-\chi)^2 + \ldots \right) r_s \ln r_s ,\qquad
\end{equation}
where as usual we neglect higher order terms in $g$ and $\chi$.

A discussion of the physically different limit of small $r_s$ and finite $g$, $\chi$, 
for which Eq.~(\ref{Qresult}) becomes applicable, is provided in next section. 

\section{DISCUSSION}\label{discussion}
By adding up the contributions from the previous sections
we obtain that the total energy per particle in the presence of spin-orbit coupling 
of the Rashba or Dresselhaus type has the following form, 
in the limit of high density and small values of $g$ and $\chi$
\begin{eqnarray} \label{final_result}
\mathcal{E}(g,&& r_s,\chi  ) \nonumber \\
= ~ &&  \mathcal{E}_0(g,r_s,\chi)  
\nonumber\\
&&  -\bigg[ 
\frac{8\sqrt{2}}{3\pi } + 
\frac{\sqrt{2}}{48\pi} 
\chi^4 
\Big( \ln \frac{\chi}{8} + \frac{23}{12}  \Big)
+ \ldots \bigg] \frac{1}{r_s} 
\nonumber \\
&&  - \bigg[ 0.385 + \frac{(g-\chi)^2}{4} +\ldots   \bigg] + \bigg[ -   \frac{2\sqrt{2}}{3\pi}  (10-3\pi) 
\nonumber \\
&& + 
\frac{56-15\pi}{40\sqrt{2}\pi}(g-\chi)^2 +\ldots  \bigg] r_s \ln r_s , \nonumber\\
\end{eqnarray}
where the non-interacting energy $\mathcal{E}_0(g,r_s,\chi)$ is given in Eq.~(\ref{nonint_energy}).
We included in Eq.~(\ref{final_result}) all the quadratic terms in $\chi$ and $g$ as well as, for the exchange energy only (third line), the term of order $\mathcal{O}(\chi^4)$ which represents in practice the leading correction at high densities. The fourth line shows the contribution of the
direct and exchange second order diagrams, from Eqs.~(\ref{E2D_final}) and (\ref{E2X_final}).
Finally, the last term (in the fourth and fifth lines) represents the ring diagrams correlation energy. 
For $g=\chi=0$ this expression recovers the classic result for the two-dimensional electron liquid of 
Rajagopal and Kimball of Ref.~\onlinecite{rajagopal77}.

From Eq.~(\ref{final_result}) it becomes clear that, for small $r_s$ and given spin-orbit coupling $\bar\alpha$, 
the largest effect of the interaction is due to the exchange energy. In particular, we could obtain in Eq.~(\ref{deltachi}) how 
the best possible repopulation, i.e., the value of $\chi$ in equilibrium, is modified by the interactions to leading 
order. We also determined in Eq.~(\ref{deltaExcResult}) the leading correction of the 
exchange-correlation energy due to the spin-orbit coupling. In agreement with the general property discussed 
in Sec.~\ref{diagrams_Section}, such corrections are very small. Beyond the perturbative regime, we
have shown (see Fig.~\ref{MC_figure}) that recent quantum Monte Carlo data are well reproduced by neglecting the correction to the
exchange-correlation energy due to the spin-orbit coupling. Our results lend some measure of comfort to the otherwise 
uncontrolled procedure of making use of spin-orbit coupling free exchange and correlation functionals within density 
functional calculations.\cite{valinrodriguez02a,valinrodriguez02b} 

We end our discussion with an analysis of the formally interesting case of the limit of
small $r_s$ at finite $g$ and $\chi$. This corresponds to the case of a diverging bare spin-orbit
coupling constant $\bar \alpha$ [see Eq.~(\ref{gdef})], a scenario that does not apply to the plain two dimensional electron liquid treated
up to this point. In this case the otherwise small `anomalous' integration region  
contributing to Eq.~(\ref{rings_sum}) becomes dominant and Eq.~(\ref{Qresult}) is the appropriate form for $Q_q(u)$. 
While the reader is referred to Appendix \ref{different_limit} for the details of the calculation, 
we give here the result for the correlation energy to quadratic order in $g$ and $\chi$
\begin{eqnarray}\label{ring_small}
\delta \mathcal{E}_R(g,r_s,\chi)\simeq -\frac{r_s \ln r_s}{ 6\sqrt{2}\pi} 
\bigg[
(\frac{208}{5}-\frac{51 \pi}{4}) \, g^2 
\nonumber \\
-  (\frac{42}{5}-\frac{9 \pi}{4}) \, \chi^2 + \frac{27}{10} (-16 + 5 \pi) \, g  \, \chi
\bigg] . 
\end{eqnarray}
This expression is distinct from Eq.~(\ref{delta_ER_normal}) and violates the general form $\propto(\chi-g)^2$. As a consequence, 
the quadratic term in the spin-orbit coupling survives and the correction is proportional to $g^2 r_s \ln r_s$ instead of $g^4 r_s \ln r_s$ (using $\chi \simeq g$). 
For the exchange energy and the second-order correlations the non-analyticity in 
the $r_s,g\to 0$ limit is not present, and the same results obtained before are valid here. Despite this fact, the ring-diagrams contribution remains a subleading correction since Eq.~(\ref{ring_small}) is applicable only when $g \gg r_s$ which
clearly implies $g^4\ln g/r_s \gg g^2 r_s \ln r_s $, i.e., the exchange energy correction is larger.

As a final remark, we notice that the nonanalyticity of the $r_s,g  \to 0$ limit discussed here becomes relevant in gated heterostructures since the increase of the density is naturally accompanied
by a modification of the confining potential. If smaller values of $r_s$ require higher values of the external electric field (e.g., proportional to the electron density\cite{WinklerSpringer03}), larger values of the spin-orbit coupling $\bar\alpha$ are obtained at the same time, thereby making the limit of Eq.~(\ref{ring_small}) meaningful. Furthermore, in heterostructures with a back gate, $r_s$ and $\bar\alpha$ can be controlled separately.

\begin{acknowledgments}
SC acknowledges support by NCCR Nanoscience, Swiss NSF, and CIFAR.

\end{acknowledgments}

\appendix

\section{Derivation of Eqs.~(\ref{E2D_final}) and (\ref{E2X_final})} \label{deriv2direct}

We consider here the second derivatives of $\mathcal{E}_2^D(g,\chi)$ and $\mathcal{E}_2^X(g,\chi)$, given by
Eqs.~(\ref{f2Ddef}) and (\ref{f2Xdef}), respectively. We start by listing some useful formulas for spin summations: 
\begin{eqnarray}
\label{spinsum00}
&&\sum_{\{\mu_i\}} \prod_{i=1}^{N}
\langle {\bf p}_i \mu_i | {\bf p}_{i+1} \mu_{i+1} \rangle = 2 ,\\
\label{spinsum22}
&&\sum_{\{\mu_i\}} \mu_j \mu_k \prod_{i=1}^{N}
\langle {\bf p}_i \mu_i | {\bf p}_{i+1} \mu_{i+1} \rangle =
2\frac{{\bf p}_j \cdot {\bf p}_k}{p_j p_k},
\end{eqnarray}
from which the following three identities are obtained
\begin{eqnarray}
\label{spinsum1}
&&\sum_{\mu,\nu}\mu (\nu p-\mu k)|\langle {\bf p}\nu |{\bf k}\mu \rangle|^2=\frac{2{\bf k}\cdot{\bf q}}{k},\\
\label{spinsum2}
&&\sum_{\mu,\mu',\nu,\nu'}
|\langle {\bf p} \, \nu | {\bf k} \mu \rangle|^2 \, 
|\langle {\bf p}'\, \nu' | {\bf k}' \mu' \rangle |^2 \nonumber\\
&&\hspace{1cm}\times (\nu p-\mu k+\nu' p'-\mu' k')^2 = 8 q^2 .\\
\label{spinsum4}
&&\sum_{\mu,\mu',\nu,\nu'}
\langle {\bf p} \, \nu | {\bf k} \mu \rangle \, 
\langle {\bf k} \mu | {\bf p}' \nu ' \rangle 
\langle {\bf p}'  \nu ' | {\bf k}' \mu ' \rangle \,
\langle {\bf k}' \mu' | {\bf p} \, \nu \rangle \nonumber \\
&&\hspace{1cm}\times (\nu p-\mu k+\nu' p'-\mu' k')^2 = 0 , 
\end{eqnarray}
where, as in Eqs.~(\ref{f2Ddef}) and (\ref{f2Xdef}), ${\bf p}={\bf k}+{\bf q}$ and ${\bf p}'={\bf k}'-{\bf q}$.

We now examine $\left. \frac{\partial^2 \mathcal{E}_2^D}{\partial g^2}\right|_0$. After calculating the second derivative of the integrand, which only involves the energy denominator, the spin summation can be evaluated by making use of (\ref{spinsum2}). One can next evaluate the angular integration in the $d{\bf q}$ integral, and than integrate in $dk_y$, $dk'_y$ to obtain the following expression
\begin{equation}\label{derivD_gg}
\left.
\frac{\partial^2 \mathcal{E}_2^D}{\partial g^2}
\right|_0 = 
-\frac{1}{\pi^2}
\int_{0}^{\infty} \frac{d q}{q^2}
\int_{-1}^1  d{k_x}
d{k'_x} \,
\frac{L(q,k_x)L(q,-k'_x)}{(q+k_x-k'_x)^3} .
\end{equation}
Here $L(q, k_x)=\int n_0(k)(1-n_0(p)) \, dk_y$, 
which in the  interval $-1\leq k_x \leq 1$ gives
\begin{equation}\label{Ldef}
L(q,k_x)=
\left\{
\begin{array}{cl}
0					        &~{\rm if}~q<2~{\rm and}~k_x \leq -\frac{q}{2}  ,\\
\ell(k_x) &  ~{\rm if}~k_x \geq -q+1 , \\
\ell(k_x)-\ell(k_x+q) & ~{\rm otherwise}  .
\end{array}
\right. 
\end{equation}
where we defined $\ell(k_x)=2\sqrt{1-k_x^2}$.
We finally have evaluated the integral (\ref{derivD_gg}) and obtained a result numerically equal
to $-\frac12$. The remaining second derivatives of $\mathcal{E}_2^D$ can be obtained from the general relation
Eq.~(\ref{derivatives_rel_D}) but, as an example, we calculate them here explicitly.

The expression of $\left.\frac{\partial^2 \mathcal{E}_2^D}{\partial g\partial \chi}\right|_0$
can be simplified by making use of (\ref{spinsum00}) and (\ref{spinsum22}). After some 
further manipulation we obtain
\begin{eqnarray}\label{derivD_cg}
\left.
\frac{\partial^2 \mathcal{E}_2^D}{\partial \chi \partial g}
\right|_0
= \,
-\frac{1}{4\pi^3}
\int \frac{d{\bf q}}{q^2}
\int d{\bf k}  
\int  d{\bf k}' \,
\frac{
n_0(k')(1-n_0(p'))}{
(q^2+{\bf q} \cdot({\bf k}-{\bf k}'))^2} \nonumber \\
\times \left[
({\bf k}\cdot {\bf q})  \,
(1-n_0(p)) \, \delta(1-k) 
-({\bf p}\cdot {\bf q}) \,
n_0(k) \, \delta(1-p) 
\right] .\nonumber\\
\end{eqnarray}
Then, after a change of variable ${\bf k}\to -{\bf k}-{\bf q}=-{\bf p}$ in
the second term in the integrand, angular integration in the $d{\bf q}$
integral, and integration in $dk_y$ and $dk'_y$, we obtain
\begin{eqnarray}\label{derivD_gc2}
\left.
\frac{\partial^2 \mathcal{E}_2^D}{\partial \chi \partial g}
\right|_0=&&
-\frac{1}{\pi^2} 
\int_{0}^{\infty} \frac{d q}{q^2}
\int_{-1}^1  d{k_x}
d{k'_x} \,
\frac{k_x \, L(q,-k_x')}{\sqrt{1-k_x^2}}  \nonumber \\
&& \times\left[
\frac{1-n_0(p)}{(q+ k_x- k'_x)^2}+
\frac{n_0(p)}{(k_x+k'_x)^2}
\right], \quad
\end{eqnarray}
where $p=\sqrt{1+q^2+2qk_x}$. Finally, Eq.~(\ref{derivD_gc2}) can be transformed to
the opposite of (\ref{derivD_gg}), by means of an integration 
by parts in $dk_x$ of the two terms of the integrand and and a suitable change 
of variable $k_x \to -k_x-q$ in the second one. 

The last term is $\left. \frac{\partial^2 \mathcal{E}_2^D}{\partial \chi^2}\right|_0$. 
It is convenient to transform terms containing double derivatives of
the occupation functions in the following way
\begin{equation}\label{partint}
\int f({\bf k})\left. \frac{\partial^2 n_\mu(k)}{\partial\chi^2} \right|_0 \,
d{\bf k}=
\frac{1}{4} \int 
\frac{\partial f({\bf k})}{\partial k}
\delta(1-k) \,
d{\bf k} .
\end{equation}
After then performing the spin summations explicitly, a change of primed and unprimed variables 
to collect similar terms, and the angular integration in the ${d}{\bf q}$ integral, we obtain
\begin{eqnarray}\label{derivD_cc}
&&
\left.
\frac{\partial^2 \mathcal{E}_2^D}{\partial \chi^2}
\right|_0 = 
-\frac{1}{2\pi^2} 
\int_0^\infty \frac{dq}{q^2}
 \int d{\bf k} 
\int d{\bf k}'  \,
n_0(k')(1-n_0(p'))\nonumber \\
&& \times\left[ 
\left(
\frac{\partial}{\partial k}~ 
\frac{1-n_0(p)}{q+k_x-k'_x} 
\right) 
\delta(1-k)
\right. 
-\left(
\frac{\partial}{\partial p}~
\frac{n_0(k)}{q+k_x-k'_x}
\right) \nonumber \\
&& ~ \times \delta(1-p)
- \frac{
2 \cos(\phi_{\bf k}-\phi_{\bf p})}
{q+k_x-k'_x} ~ \delta(1-k) \delta(1-p) 
\bigg] , 
\end{eqnarray}
where in the $d{\bf k}$, $d{\bf k}'$ integrations, the $x$ axis is 
chosen to be along ${\bf q}$ so that $p=\sqrt{k^2+q^2+2q k_x }$.
It is also convenient to change variable ${\bf k} \to -{\bf k}-{\bf q}$ in the 
second term in the square brackets. 
The contributions from the first two terms in the square brackets involving the 
derivative $\frac{\partial n_0(p)}{\partial k}= -\delta(1-p)\frac{\partial p}{\partial k}$ 
cancel exactly the third term. In fact, we obtain that the coefficient, 
multiplying the product $\delta(1-k) \delta(1-p)$ has the following form
\begin{equation}\label{deltas_coeff}
\frac{\partial p/\partial k}{q+k_x-k'_x}
-\frac{\partial p/\partial k}{k_x+k'_x}
-\frac{2\cos(\phi_{\bf k}-\phi_{\bf p})}{q+k_x-k'_x} ,
\end{equation}
which can be simplified using $k_x=\cos\phi_{\bf k}=-q/2$.
Furthermore, $\partial p/\partial k=(k+ q \cos\phi_{\bf k})/p=1-q^2/2$ and
$\cos(\phi_{\bf k}-\phi_{\bf p})=1-q^2/2$ as well. Therefore 
(\ref{deltas_coeff}) is seen to vanish identically.  The term surviving
in Eq.~(\ref{derivD_cc}), after integrations in $dk_y$, $dk'_y$, is
the opposite of (\ref{derivD_gc2}). Therefore, we can summarize the final result as follows:
\begin{equation}\label{frac12}
\left.\frac{\partial^2 \mathcal{E}_2^D}{\partial g^2}\right|_0 =
-\left.\frac{\partial^2 \mathcal{E}_2^D}{\partial \chi \partial g}\right|_0 =
\left.\frac{\partial^2 \mathcal{E}_2^D}{\partial \chi^2}\right|_0
= -\frac12.
\end{equation}

We turn now to the second derivatives of $\mathcal{E}_2^X$. The fact that $\left.\frac{\partial^2 \mathcal{E}_2^X}{\partial g^2} \right|_0=0$
is immediately obtained from the spin summation (\ref{spinsum4}) and the mixed derivative is also found to vanish upon spin summation. Finally, the 
second derivative with respect to $\chi$ is found to be vanishing after rather 
cumbersome manipulations. These results are consistent with the general property 
(\ref{derivatives_rel_D}) and we conclude that:
\begin{equation}\label{zero}
\left.\frac{\partial^2 \mathcal{E}_2^X}{\partial g^2}\right|_0 =
-\left.\frac{\partial^2 \mathcal{E}_2^X}{\partial \chi \partial g}\right|_0 =
\left.\frac{\partial^2 \mathcal{E}_2^X}{\partial \chi^2}\right|_0
= 0.
\end{equation}

\section{Second derivatives of the ring diagrams contribution}  \label{deriv2ringdiagrams}

We consider here the second derivatives with respect to $g$ and $\chi$ of $Q_q(u)$, defined in Eq.~(\ref{Qdef}). 
We begin by noting that
\begin{eqnarray}\label{ddQ_gg_final}
\left. \frac{\partial^2 Q_q(u)}{\partial g^2}\right|_0=16 \int 
\frac{d{\bf k}}{q}n_0(k)(1-n_0(p))(q+2k_x)\nonumber \\
\times\frac{(q+2k_x)^2-12 u^2}{[(q+2k_x)^2+4u^2]^3}.\hspace{1cm}
\end{eqnarray}
where we have set $\chi=g=0$ and performed the spin summation (which gives a factor of 2). Consider now the mixed derivative
\begin{eqnarray}
\left. \frac{\partial^2 Q_q(u)}{\partial \chi\partial g}\right|_0=\int \frac{d{\bf k}}{q^2}\sum_{\mu,\nu}\left[\frac{\mu\delta(1-k)}{2}(1-n_0(p))\right. \nonumber \\
\left. -\frac{\nu\delta(1-p)}{2}n_0(k)\right](\nu p-\mu k)|\langle {\bf p}\nu |{\bf k}\mu \rangle|^2\nonumber \\
\times\frac{4[(q+2k_x)^2-4 u^2]}{[(q+2k_x)^2+4u^2]^2}.\hspace{2.5cm}
\end{eqnarray}
By change of variable ${\bf k}\to -{\bf k}-{\bf q}$ and relabeling $\mu\leftrightarrow \nu $ in the second term of the integrand (from the large square parenthesis), we can cancel the $n_0(p)$ contribution in the first term. Finally, the spin summation can be performed by using Eq.~(\ref{spinsum1}) and the integration in $d{\bf k}$ gives
\begin{equation}\label{ddQ_chi_g_final}
\left. \frac{\partial^2 Q_q(u)}{\partial g \partial \chi}\right|_0=\int_0^{2\pi} \frac{4 \cos{\phi}}{q}\frac{(q+2\cos\phi)^2-4 u^2}{[(q+2\cos\phi)^2+4 u^2]^2}\,d\phi .
\end{equation}

The last second derivative, with respect to $\chi$, gives
\begin{eqnarray}
\left. \frac{\partial^2 Q_q(u)}{\partial \chi^2}\right|_0=\int \frac{d{\bf k}}{q}\left[\frac{\delta'(1-k)-\delta(1-k)}{4}(1-n_0(p))\right. \nonumber \\
\left. -\frac{\delta'(1-p)-\delta(1-p)}{4}n_0(k)-\frac{\delta(1-k)\delta(1-p)}{2}\right]\nonumber \\
\times\frac{8(q+2k_x)}{(q+2k_x)^2+4u^2},\hspace{4cm}
\end{eqnarray}
The derivatives only involve $n_\mu(k)(1-n_\nu(p))$, which results in the square parenthesis. The third term in the integrand is vanishing, since $p=q=1$ implies $q+2k_x=0$ (note that ${\bf p}={\bf k}+{\bf q}$ and ${\bf q}$ is along $x$).
Furthermore, we can change variable in the second term and cancel the $n_0(p)$ contribution of the first term. Therefore, the square parenthesis simplifies to $[\delta'(1-k)-\delta(1-k)]/4$ and the integration in $dk$ is immediate. The final result coincides with the opposite of Eq.~(\ref{ddQ_chi_g_final}). Consider now the change of variable $\cos\phi \to k_x$, which gives
\begin{eqnarray}\label{ddQ_cc_final}
&&\left. \frac{\partial^2 Q_q(u)}{\partial \chi^2}\right|_0=\int_{-1}^{1} \frac{8 k_x dk_x}{q\sqrt{1-k_x^2}}
\frac{(q+2k_x)^2-4 u^2}{[(q+2k_x)^2+4 u^2]^2}\nonumber \\
&=&-\int_{-1}^{1} \frac{8 dk_x}{q}\sqrt{1-k_x^2} \frac{\partial}{\partial k_x}
\frac{(q+2k_x)^2-4 u^2}{[(q+2k_x)^2+4 u^2]^2}\nonumber \\
&=&32 \int_{-1}^{1} 
\frac{dk_x}{q}\sqrt{1-k_x^2}(q+2k_x)\frac{(q+2k_x)^2-12 u^2}{[(q+2k_x)^2+4u^2]^3},\nonumber \\
\end{eqnarray}
where we integrated by parts in the second line. Notice that for $q \geq 2$
(see Appendix \ref{deriv2direct})
$\int dk_y  n_0(k)(1-n_0(p))=2\sqrt{1-k_x^2}$. This establishes the 
equivalence of Eq.~(\ref{ddQ_cc_final}) and (\ref{ddQ_gg_final}) for this case. 
The equivalence of the two expressions holds also at $q<2$, as can be seen applying the 
change of variable $k_x \to -k_x-q$ in the integration region 
$-1<k_x<-\frac{q}{2}$ of (\ref{ddQ_cc_final}). Therefore, we conclude that
\begin{equation}\label{ddQ_chi_final}
Q^{gg}_q(u)=\left. \frac{\partial^2 Q_q(u)}{\partial g^2}\right|_0=-\left. 
\frac{\partial^2 Q_q(u)}{\partial \chi\partial g}\right|_0 = \left. 
\frac{\partial^2 Q_q(u)}{\partial \chi^2}\right|_0,
\end{equation}
This is in agreement with Eq.~(\ref{derivatives_rel_D}) and leads to (\ref{Q_right_form}).

\section{Physically alternate limit $r_s\to 0$ for finite $g$, $\chi$} \label{different_limit}
In this Appendix we analyze the $r_s\to 0$ limit of Eq.~(\ref{rings_sum}) for fixed values of 
$g$ and $\chi$. In this situation the relevant integration region in $dq$
is of order $r_s\ll g,\chi$ around $q=0$, and an `anomalous' quadratic correction in $g$ to 
the final result for the energy is obtained [see the discussion after Eq.~(\ref{Q_right_form})].
The calculation can be patterned after that carried out in the absence 
of spin-orbit coupling, as for instance done in Ref.~\onlinecite{rajagopal77}. 
In this case one can perform the integration in $d q$ (up to an arbitrary upper 
limit much larger than $r_s$) and extracts the coefficient of the $r_s \ln r_s$ contribution
by writing
\begin{equation}\label{ring_approx}
\mathcal{E}_R(g,r_s,\chi)\simeq -
\frac{r_s \ln{r_s}}{ 3\sqrt{2}(2\pi)^4}  \int_{-\infty}^{+\infty} [Q_0(u)]^3 \, du  ~,
\end{equation}
which gives the standard result $-\frac{2\sqrt{2}}{3\pi}(10-3\pi)r_s\ln r_s$ by using
$Q_0(u)=4\pi R(u)$. Using for $Q_0(u)$ the expression of Eq.~(\ref{Qresult}) instead, 
the result can still be obtained analytically as in Ref.~\onlinecite{chesi07a}, i.e.
\begin{eqnarray}\label{ring_result}
\mathcal{E}_R(g,r_s,\chi)\simeq -\frac{r_s \ln r_s}{ 6\sqrt{2}\pi} 
\left[
(10-3\pi)
\left(
\frac{k_+^3}{\tilde k_+^2}+
\frac{k_-^3}{\tilde k_-^2}
\right) \right . \quad\\
\left.
+ \frac{3 k_+ k_-^2}{\tilde k_+ \tilde k_-^2} 
F(\tilde k_+ ,\tilde k_-)
+  \frac{3 k_- k_+^2}{\tilde k_- \tilde k_+^2} 
F(\tilde k_- ,\tilde k_+)
\right] ~, \nonumber
\end{eqnarray}
where we have defined the following function:
\begin{equation}\label{def_F}
F(x,y)=4(x+y)-\pi x
-4 x E(1-\frac{y^2}{x^2})
+\frac{2 x^2 \arccos\frac{y}{x}}{\sqrt{(x-y)(x+y)}}~,
\end{equation}
in terms of the elliptic function $E(x)$, defined as in Ref.~\onlinecite{Abramowitz65}.
Notice that one can use the identity $\frac{\arccos\frac{y}{x}}{\sqrt{x-y}}=
\frac{{\rm arccosh}\frac{y}{x}}{\sqrt{y-x}}$ when $y>x$. 


\begin{thebibliography}{50}%

\bibitem [{ras()}]{rashbaSO}
Y. A. Bychkov and E. I. Rashba, JETP Lett. {\bf 39}, 78 (1984);
J. Phys. C {\bf 17}, 6039 (1984).
\bibitem{datta89}
S. Datta and B. Das, Appl. Phys. Lett. {\bf 56}, 665 (1989).
\bibitem{wolf01}
S. A. Wolf, D. D. Awschalom, R. A. Buhrman, J. M. Daughton, S. von Molnar, M. L. Roukes, A. Y. Chtchelkanova, and D. M. Treger, Science {\bf 294}, 1488 (2001).
\bibitem{zutic04}
I. \v{Z}uti\'c, J. Fabian, and S. Das Sarma, Rev. Mod. Phys. {\bf 76}, 323 (2004).
\bibitem{rokhinson04}
L. P. Rokhinson, V. Larkina, Y. B. Lyanda-Geller, L. N. Pfeiffer, and K. W. West, Phys. Rev. Lett. {\bf 93}, 146601 (2004).
\bibitem{khodas04}
M. Khodas, A. Shekhter, and A. M. Finkel\'stein, Phys. Rev. Lett. {\bf 92}, 086602 (2004).
\bibitem{chen05}
H. Chen, J. J. Heremans, J. A. Peters, A. O. Govorov, N. Goel, S. J. Chung, and M. B. Santos, Appl. Phys. Lett. {\bf 86}, 032113 (2005).
\bibitem{meier07}
L. Meier, G. Salis, I. Shorubalko, E. Gini, S. Sch\"on, and K. Ensslin, Nat. Phys. {\bf 3}, 650 (2007).
\bibitem{chesi07c}
S. Chesi and G. F. Giuliani, cond-mat/0701415.
\bibitem{Chesi2011qpc}
S. Chesi, G. F. Giuliani, L. P. Rokhinson, L. N. Pfeiffer, and K. W. West, arXiv:1011.2676.
\bibitem{TheBook}
G. F. Giuliani and G. Vignale, \emph{Quantum Theory of the Electron Liquid} (Cambridge University Press, Cambridge, 2005).
\bibitem{WinklerSpringer03}
R. Winkler, \emph{Spin-Orbit Coupling Effects in Two-Dimensional Electron and Hole Systems} (Springer, Berlin, 2003).
\bibitem{winkler05}
R. Winkler, E. Tutuc, S. J. Papadakis, S. Melinte, M. Shayegan, D. Wasserman, and S. A. Lyon, Phys. Rev. B {\bf 72}, 195321 (2005).
\bibitem{Chesi2007exchange}
S. Chesi and G. F. Giuliani, Phys. Rev. B {\bf 75}, 155305
(2007).
\bibitem{Chen1999}
G.-H. Chen and M. E. Raikh, Phys. Rev. B {\bf 60}, 4826 (1999).
\bibitem{Saraga2005}
D. S. Saraga and D. Loss, Phys. Rev. B {\bf 72}, 195319 (2005).
\bibitem{Giuliani2005}
G. F. Giuliani and S. Chesi, in \emph{Highlights in the Quantum Theory of Condensed Matter}, ed. F. Beltram, p. 269
(Edizioni della Normale, Pisa, 2005).
\bibitem{Wang2005}
X. F. Wang, Phys. Rev. B {\bf  72}, 085317 (2005). 
\bibitem{Dimitrova2005}
O. V. Dimitrova, Phys. Rev. B {\bf 71}, 245327 (2005).
\bibitem{Schliemann2006}
J. Schliemann, Phys. Rev. B {\bf 74}, 045214 (2006). 
\bibitem{Pletyukhov2006}
M. Pletyukhov and V. Gritsev, Phys. Rev. B {\bf 74}, 045307 (2006). 
\bibitem{Chesi2007phasediagr} 
S. Chesi, G. Simion, and G. F. Giuliani, cond-mat/0702060.
\bibitem{ChesiPhD}
S. Chesi, Ph.D. thesis, Purdue University, 2007.
\bibitem{Badalyan2009}
S. M. Badalyan, A. Matos-Abiague, G. Vignale, and J. Fabian, Phys. Rev. B {\bf 79}, 205305 (2009).
\bibitem{Ambrosetti2009}
A. Ambrosetti, F. Pederiva, E. Lipparini, and S. Gandolfi, Phys. Rev. B {\bf 80}, 125306 (2009).
\bibitem{Nechaev2009} 
I. A. Nechaev, M. F. Jensen, E. D. L. Rienks, V. M. Silkin, P. M. Echenique, E. V. Chulkov, and P. Hofmann, Phys. Rev. B {\bf 80}, 113402 (2009).
\bibitem{Nechaev2010}
I. A. Nechaev, P. M. Echenique, and E. V. Chulkov, Phys. Rev. B {\bf 81}, 195112 (2010).
\bibitem{Chesi2010exact}
S. Chesi and G. F. Giuliani, arXiv:1008.2227. 
\bibitem{Abedinpour2010}
S. H. Abedinpour, G. Vignale, and I. V. Tokatly, Phys. Rev. B {\bf 81}, 125123 (2010). 
\bibitem{Zak2010}
R. A. \.Zak, D. L. Maslov, and D. Loss, Phys. Rev. B {\bf 82}, 115415 (2010).
\bibitem{Schliemann2010}
J. Schliemann, Europhys. Lett. {\bf 91}, 67004 (2010). 
\bibitem{Agarwal2011}
A. Agarwal, S. Chesi, T. Jungwirth, J. Sinova, G. Vignale, and M. Polini, Phys. Rev. B {\bf 83}, 115135 (2011).
\bibitem{dresselhaus55}
G. Dresselhaus, Phys. Rev. {\bf 100}, 580 (1955).
\bibitem{comment_dresselhaus}
More specifically, we refer here and in the rest of the paper 
to the linear Dresselhaus spin-orbit coupling which appears in quantum wells with 
confinement along [001].
\bibitem{rajagopal77}
A. K. Rajagopal and J. C. Kimball, Phys. Rev. B {\bf 15}, 2819 (1977).
\bibitem{gellmann57}
M. Gell-Mann and K. A. Brueckner, Phys. Rev. {\bf 106}, 364 (1957).
\bibitem{chesi07a}
S. Chesi and G. F. Giuliani, Phys. Rev. B {\bf 75}, 153306 (2007).
\bibitem {attaccalite02}
C. Attaccalite, S. Moroni, P. Gori-Giorgi, and G. B. Bachelet, Phys. Rev. Lett. {\bf 88}, 256601 (2002).
\bibitem{valinrodriguez02a}
M. Val\'in-Rodr\'iguez, A. Puente, and L. Serra, Phys. Rev. B {\bf 66}, 045317 (2002).
\bibitem{valinrodriguez02b}
M. Val\'in-Rodr\'iguez, A. Puente, L. Serra, and E. Lipparini, Phys. Rev. B {\bf 66}, 165302 (2002).
\bibitem{comment_no_bbroblem}
 Notice that the function $Q_q(u)$ does not
  have to satisfy condition Eq.~(\ref{diagr_generic_form}); the energy does. On
  the other hand, if $Q_q(u)$ does satisfy Eq.~(\ref{diagr_generic_form}), one
  can immediately conclude that the corresponding energy correction is behaving
  properly.
\bibitem{Abramowitz65}
M. Abramowitz and I. A. Stegun, \emph{Handbook of Mathematical Functions} (Dover Publications, New York, 1965).

\end{thebibliography}

%

\end{document}